\documentclass[]{spie}  

\usepackage{amsmath,amsfonts,amssymb}
\usepackage{graphicx}
\usepackage[colorlinks=true, allcolors=blue]{hyperref}
\usepackage{booktabs}

\usepackage{silence}
\usepackage{siunitx}
\usepackage{caption}
\usepackage{subcaption}
\usepackage{orcidlink}
\usepackage{microtype}
\usepackage{xspace}

\newcommand{\BK}{BICEP\xspace}
\newcommand{\TtoP}{$T\!\rightarrow\!P$\xspace}
\newcommand{\TtoB}{$T\!\rightarrow\!B$\xspace}
\newcommand{\Dr}{\Delta(r)}

\title{Quantifying the systematic impact of differential beam response on the
  \BK CMB polarization data from 2016 through 2024}

\author[a,b]{\href{https://orcid.org/0000-0003-4117-6822}{B.~D.~Elwood}}%
\author[c]{P.~A.~R.~Ade}%
\author[d,e]{\href{https://orcid.org/0000-0002-9957-448X}{Z.~Ahmed}}%
\author[f]{\href{https://orcid.org/0000-0001-6523-9029}{M.~Amiri}}%
\author[a]{\href{https://orcid.org/0000-0002-8971-1954}{D.~Barkats}}%
\author[g]{\href{https://orcid.org/0000-0002-3351-3078}{R.~Basu~Thakur}}%
\author[h]{\href{https://orcid.org/0000-0001-9185-6514}{C.~A.~Bischoff}}%
\author[i]{\href{https://orcid.org/0000-0003-0848-2756}{D.~Beck}}%
\author[g,j]{J.~J.~Bock}%
\author[k]{V.~Buza}%
\author[i,d]{\href{https://orcid.org/0000-0003-4541-7080}{B.~Cantrall}}%
\author[g]{\href{https://orcid.org/0000-0002-1630-7854}{J.~R.~Cheshire~IV}}%
\author[l]{J.~Connors}%
\author[m]{\href{https://orcid.org/0000-0002-2088-7345}{J.~Cornelison}}%
\author[n]{M.~Crumrine}%
\author[g]{A.~J.~Cukierman}%
\author[l]{E.~Denison}%
\author[o]{L.~Duband}%
\author[a]{\href{https://orcid.org/0000-0002-7059-8728}{M.~A.~Echter}}%
\author[p]{\href{https://orcid.org/0009-0007-6718-1730}{M.~Eiben}}%
\author[g]{\href{https://orcid.org/0000-0002-3790-7314}{S.~Fatigoni}}%
\author[q]{\href{https://orcid.org/0000-0001-8217-6832}{J.~P.~Filippini}}%
\author[i]{A.~Fortes}%
\author[g]{M.~Gao}%
\author[h]{C.~Giannakopoulos}%
\author[i]{N.~Goeckner-Wald}%
\author[i]{\href{https://orcid.org/0000-0001-5268-8423}{D.~C.~Goldfinger}}%
\author[r,s]{S.~Gratton}%
\author[i]{J.~A.~Grayson}%
\author[g]{\href{https://orcid.org/0009-0003-6999-0129}{A.~Greathouse}}%
\author[a]{\href{https://orcid.org/0000-0001-9292-6297}{P.~K.~Grimes}}%
\author[f]{M.~Halpern}%
\author[d,e]{S.~Henderson}%
\author[n]{\href{https://orcid.org/0000-0002-3437-5228}{T.~D.~Hoang}}%
\author[l]{J.~Hubmayr}%
\author[g]{\href{https://orcid.org/0000-0001-5812-1903}{H.~Hui}}%
\author[i]{K.~D.~Irwin}%
\author[t]{M.~Izquierdo~Poza}%
\author[g]{\href{https://orcid.org/0000-0002-3470-2954}{J.~H.~Kang}}%
\author[t]{\href{https://orcid.org/0000-0002-5215-6993}{K.~S.~Karkare}}%
\author[g]{S.~Kefeli}%
\author[a,b]{\href{https://orcid.org/0009-0003-5432-7180}{J.~M.~Kovac}}%
\author[i]{C.~Kuo}%
\author[n,u]{\href{https://orcid.org/0000-0002-4540-1495}{K.~Lasko}}%
\author[g]{\href{https://orcid.org/0000-0002-6445-2407}{K.~Lau}}%
\author[h]{M.~Lautzenhiser}%
\author[i]{\href{https://orcid.org/0000-0001-5677-5188}{T.~Liu}}%
\author[k,v]{\href{https://orcid.org/0000-0002-1414-7236}{S.~C.~Mackey}}%
\author[n]{N.~Maher}%
\author[j]{K.~G.~Megerian}%
\author[g]{L.~Minutolo}%
\author[g]{\href{https://orcid.org/0000-0002-4242-3015}{L.~Moncelsi}}%
\author[i]{Y.~Nakato}%
\author[g,j]{H.~T.~Nguyen}%
\author[g,j]{R.~O’Brient}%
\author[a]{S.~N.~Paine}%
\author[g]{A.~Patel}%
\author[a]{\href{https://orcid.org/0000-0002-4436-4215}{M.~A.~Petroff}}%
\author[a,b]{\href{https://orcid.org/0000-0002-7822-6179}{A.~R.~Polish}}%
\author[o]{T.~Prouve}%
\author[n]{\href{https://orcid.org/0000-0003-3983-6668}{C.~Pryke}}%
\author[l]{C.~D.~Reintsema}%
\author[g]{T.~Romand}%
\author[i]{M.~Salatino}%
\author[g]{A.~Schillaci}%
\author[a]{B.~Schmitt}%
\author[n,u]{\href{https://orcid.org/0000-0001-7387-0881}{B.~Singari}}%
\author[g,j]{A.~Soliman}%
\author[a]{T.~St.~Germaine}%
\author[g]{\href{https://orcid.org/0000-0003-0260-605X}{A.~Steiger}}%
\author[g]{B.~Steinbach}%
\author[c]{R.~Sudiwala}%
\author[i,d]{K.~L.~Thompson}%
\author[c]{\href{https://orcid.org/0000-0002-1851-3918}{C.~Tucker}}%
\author[j]{A.~D.~Turner}%
\author[w]{\href{https://orcid.org/0000-0002-3942-1609}{C.~Verg\`{e}s}}%
\author[k,v]{A.~G.~Vieregg}%
\author[g]{\href{https://orcid.org/0000-0002-8232-7343}{A.~Wandui}}%
\author[j]{A.~C.~Weber}%
\author[n]{\href{https://orcid.org/0000-0002-6452-4693}{J.~Willmert}}%
\author[g,d,e]{\href{https://orcid.org/0000-0001-5411-6920}{W.~L.~K.~Wu}}%
\author[i]{H.~Yang}%
\author[k,m]{\href{https://orcid.org/0000-0002-8542-232X}{C.~Yu}}%
\author[a]{\href{https://orcid.org/0000-0001-6924-9072}{L.~Zeng}}%
\author[d]{\href{https://orcid.org/0000-0001-8288-5823}{C.~Zhang}}%
\author[g]{S.~Zhang}%
\affil[a]{Center for Astrophysics, Harvard \& Smithsonian, Cambridge, MA 02138, USA}%
\affil[b]{Department of Physics, Harvard University, Cambridge, MA 02138, USA}%
\affil[c]{School of Physics and Astronomy, Cardiff University, Cardiff, CF24 3AA, UK}%
\affil[d]{Kavli Institute for Particle Astrophysics and Cosmology, Stanford University, Stanford, CA 94305, USA}%
\affil[e]{SLAC National Accelerator Laboratory, Menlo Park, CA 94025, USA}%
\affil[f]{Department of Physics and Astronomy, University of British Columbia, Vancouver, British Columbia, V6T 1Z1, Canada}%
\affil[g]{Department of Physics, California Institute of Technology, Pasadena, CA 91125, USA}%
\affil[h]{Department of Physics, University of Cincinnati, Cincinnati, OH 45221, USA}%
\affil[i]{Department of Physics, Stanford University, Stanford, CA 94305, USA}%
\affil[j]{Jet Propulsion Laboratory, California Institute of Technology, Pasadena, CA 91109, USA}%
\affil[k]{Kavli Institute for Cosmological Physics, University of Chicago, Chicago, IL 60637, USA}%
\affil[l]{National Institute of Standards and Technology, Boulder, CO 80305, USA}%
\affil[m]{High-Energy Physics Division, Argonne National Laboratory, Lemont, IL, 60439, USA}%
\affil[n]{School of Physics and Astronomy, University of Minnesota, Minneapolis, MN 55455, USA}%
\affil[o]{Service des Basses Temp\'eratures, Commissariat \`a l'\'Energie Atomique, 38054 Grenoble, France}%
\affil[p]{Faculty of Physical Sciences, University of Iceland, 102 Reykjav\'ik, Iceland}%
\affil[q]{Department of Physics, University of Illinois at Urbana-Champaign, Urbana, IL 61801, USA}%
\affil[r]{Centre for Theoretical Cosmology, DAMTP, University of Cambridge, Cambridge CB3 0WA, UK}%
\affil[s]{Kavli Institute for Cosmology Cambridge, Cambridge CB3 0HA, UK}%
\affil[t]{Department of Physics, Boston University, Boston, MA 02215, USA}%
\affil[u]{Minnesota Institute for Astrophysics, University of Minnesota, Minneapolis, MN 55455, USA}%
\affil[v]{Department of Physics, University of Chicago, Chicago, IL 60637, USA}%
\affil[w]{Lawrence Berkeley National Laboratory, Berkeley, CA 94720, USA}

\authorinfo{Send correspondence to B. D. Elwood: \href{mailto:bdelwood@fas.harvard.edu}{bdelwood@fas.harvard.edu}}

\begin{document}
\maketitle

\begin{abstract}
	As cosmic microwave background (CMB) polarization experiments, including
	BICEP3, BICEP Array (BA), and future BICEP experiments, achieve ever-deeper
	polarization maps in search of primordial B-modes sourced from inflation,
	constraining instrumental systematics below statistical uncertainties becomes
	progressively more challenging. Since polarimetry in the BICEP telescopes is
	performed by pair-differencing co-located, orthogonally polarized detectors,
	differential beam response leads to temperature-to-polarization (\TtoP)
	leakage, introducing a potential systematic bias on the inferred
	tensor-to-scalar ratio $r$. To mitigate this leakage, the lowest-order beam
	mismatch modes are filtered out of the CMB polarization maps through
	deprojection; however, residual undeprojected modes remain. Motivated to
	quantify this residual contamination, we perform dedicated in situ
	far-field beam measurements of the BICEP receivers during
	austral-summer calibration campaigns. We quantify the systematic impact of the
	undeprojected residuals with a specialized set of timestream simulations based
	on the measured per-detector beams. These ``beam measurement-informed
	simulations'' yield an
	estimate of the false polarized signal sourced by the undeprojected residuals,
	which can then be propagated through the cosmological analysis to assess its
	impact on $r$. We summarize the beam measurements relevant to the BK24 data
	release and present preliminary residual-leakage results for BICEP3 at
	95\,GHz. For BICEP3 over 2016--2024, deprojecting all six standard templates together with
	readout-crosstalk templates and the radially smoothed counterparts of the
	standard six in these simulations reduces the
	equivalent-$r$ leakage amplitude from
	$\rho=(4.5\pm0.7)\times10^{-3}$ to
	$(1.12\pm0.06)\times10^{-3}$. We further describe an ongoing program to extend
	the deprojection basis beyond its historical six modes, guided by a forward
	optical model that relates candidate leakage modes to perturbations of physical
	instrument parameters.
\end{abstract}


\section{INTRODUCTION}
\label{sec:intro}

Inflation predicts a background of primordial gravitational waves that imprints a
faint, degree-scale, curl-type (B-mode) pattern on the polarization of the cosmic
microwave background (CMB). A detection of primordial B modes would provide
sufficient evidence for inflation.\cite{kamionkowski2016} These modes preserve
information from the inflationary epoch, when the energy density of the Universe
was far beyond that accessible to terrestrial accelerators. Their amplitude is
parametrized by the tensor-to-scalar ratio $r$, which constrains the energy scale
of inflation.

The \BK program is a search for these primordial B modes. It operates a series
of small-aperture refracting telescopes at the South
Pole,\cite{bicep3,bicepArray} observing a low-foreground field in several
frequency bands to separate the primordial signal from Galactic dust and
synchrotron emission. The BK18 dataset, combining \BK observations through
2018 with \textit{Planck} and WMAP, constrains $r < 0.036$ at 95\% confidence
with $\sigma(r)=0.009$.\cite{bkxiii} The forthcoming
BK24 release adds deeper BICEP3 and BICEP Array maps through 2024 and targets
$\sigma(r)\sim0.005$.

In BK24, we target $\Dr\lesssim0.001$ for each individual systematic, roughly
one fifth of the projected statistical uncertainty.\cite{verges2026}
Differential beam response is potentially a leading instrumental systematic. The \BK
receivers measure polarization by differencing the signals of co-located,
orthogonally polarized detector pairs, so a mismatch between the two angular
responses of a pair converts the unpolarized temperature sky, roughly three
orders of magnitude brighter than the degree-scale B-mode signal, into false
polarization: temperature-to-polarization (\TtoP) leakage.

Boresight (``deck'') rotation modulates instrument-fixed beam mismatch
against the sky, and deprojection---a filtering operation applied to the
timestreams before mapmaking---removes a selected high-leakage subspace
(Sec.~\ref{sec:leakage}). We determine what survives these mitigations from
the measured beams themselves. In dedicated campaigns, we map each detector beam in situ
(Sec.~\ref{sec:ffbm}). We then propagate those measurements through
beam measurement-informed simulations of the full observing and analysis pipeline
(Sec.~\ref{sec:beamsims}). The resulting residual, or ``undeprojected,''
\TtoP leakage is presented in Sec.~\ref{sec:results}.

We also consider how to improve deprojection itself by expanding its mode
basis (Sec.~\ref{sec:basis}). Every mainline analysis since BICEP2 has used
the same six low-order templates, constructed from spatial derivatives of the
beam-smoothed temperature sky and conventionally associated with gain,
pointing, beamwidth, and ellipticity mismatch.\cite{bicep2iii,bkxi} Here we
present an extended deprojection basis intended for the upcoming BK24 analysis.
It adds diffraction-scale wide-mode templates and templates for readout
crosstalk between multiplexing partners. We also describe a forward optical
model that relates the wide modes to first-order perturbations of instrument
parameters and a near-field comparison that tests an instrumental
interpretation of the wide dipole.

\section{DIFFERENTIAL BEAM RESPONSE, BORESIGHT ROTATION, AND DEPROJECTION}
\label{sec:leakage}

\subsection{Pair differencing and \texorpdfstring{\TtoP}{T-to-P} leakage}
\label{sec:leakage:pairdiff}

The \BK receivers measure polarization by differencing co-located,
orthogonally polarized detectors. Let $d_A$ and $d_B$ be the timestream
samples of such a pair $A,B$, let $T(\hat n)$, $Q(\hat n)$, and $U(\hat n)$
be the Stokes fields on the sky, and let $\psi$ be the polarization angle of
detector $A$ projected on the sky. For matched beams, the pair difference is
\begin{equation}
	d\equiv\tfrac{1}{2}(d_A-d_B)
	=Q\cos2\psi+U\sin2\psi .
	\label{eq:idealpairdiff}
\end{equation}
The $\cos2\psi$ and $\sin2\psi$ components therefore determine $Q$ and $U$.

A real detector pair measures the sky through finite, potentially mismatched
angular responses. Let $B_A$ and $B_B$ be the instrument-frame beam responses
as functions of angular offset, with the detector gains included, and define
the pair-sum and pair-difference beams
$\bar B=\tfrac{1}{2}(B_A+B_B)$ and $\Delta B=B_A-B_B$. Let
$\mathcal R_\psi$ rotate a beam from instrument to sky coordinates, so that
$\bar B_\psi=\mathcal R_\psi\bar B$ and
$\Delta B_\psi=\mathcal R_\psi\Delta B$. The pair-difference timestream sample
at sky direction $\hat n_0$ is then
\begin{equation}
	d(\hat n_0;\psi)=\tfrac{1}{2}\left[T\ast\Delta B_\psi\right](\hat n_0)+\left[Q\ast\bar B_\psi\right](\hat n_0)\cos2\psi+\left[U\ast\bar B_\psi\right](\hat n_0)\sin2\psi ,
	\label{eq:pairdiff}
\end{equation}
where $\ast$ denotes convolution on the sky. The first term is a failure of
temperature to cancel in $A-B$. It vanishes for matched beams,
$\Delta B=0$, and otherwise enters the same data used to estimate $Q$ and
$U$. This is \TtoP leakage.

Genuine polarization has an exact spin-2 dependence on $\psi$, entering the
timestream through $\cos2\psi$ and $\sin2\psi$ alone, while the angular
dependence of the leakage is set by the shape and orientation of $\Delta B$.
For the $N$ samples contributing to a sky pixel at $\hat n$, with polarization
angles $\psi_k$, define the temperature-sourced leakage values and their data
vector as
\begin{equation}
	\lambda_k(\hat n)
	=\tfrac{1}{2}\left[T\ast\Delta B_{\psi_k}\right](\hat n),
	\qquad
	\boldsymbol{\lambda}(\hat n)
	=\begin{pmatrix}\lambda_1(\hat n)&\cdots&\lambda_N(\hat n)\end{pmatrix}^{\sf T}
	\in\mathbb R^N .
	\label{eq:leakage-data}
\end{equation}
Applying the mapmaking least-squares solution to this component gives the
leakage maps\cite{bkvii}
\begin{equation}
	\begin{pmatrix}
		Q_{\rm leak} \\ U_{\rm leak}
	\end{pmatrix}\!(\hat n)
	=\left(\mathbf A^{\sf T}\mathbf W\mathbf A\right)^{-1}
	\mathbf A^{\sf T}\mathbf W\,
	\boldsymbol\lambda(\hat n),
	\qquad
	\mathbf A
	=\begin{pmatrix}
		\cos2\psi_1 & \sin2\psi_1 \\
		\vdots      & \vdots      \\
		\cos2\psi_N & \sin2\psi_N
	\end{pmatrix}
	\in\mathbb R^{N\times2},
	\label{eq:leakage-stokes}
\end{equation}
where $\mathbf W$ is the diagonal matrix of inverse-variance sample weights.
The values $Q_{\rm leak}$ and $U_{\rm leak}$ are the coordinates of the
$\mathbf W$-weighted projection of $\boldsymbol\lambda$ onto the span of the
columns of $\mathbf A$, the subspace occupied by genuine polarization.
Components orthogonal to this
subspace do not enter the maps. The map-level leakage therefore depends on
both $\Delta B$ and the orientation coverage. Boresight rotation changes which
components project onto that subspace (Sec.~\ref{sec:leakage:deck}), while
deprojection subtracts selected components before $Q$ and $U$ are formed
(Sec.~\ref{sec:leakage:depj}). The simulations of Sec.~\ref{sec:beamsims} use
the real orientation coverage for this reason.

\subsection{Boresight rotation}
\label{sec:leakage:deck}

Boresight rotation suppresses leakage from beam mismatch. The compact, on-axis
\BK design permits rotation of the entire receiver about its boresight, so
that a beam imperfection fixed in the instrument frame enters the sky map at a
different orientation in each observation. The $\cos2\psi,\sin2\psi$ fit above
projects the resulting boresight-angle dependence onto the polarization
subspace. A pair-difference beam mode of azimuthal order $m$, that is with
$m$-fold azimuthal symmetry, acquires an $e^{im\psi}$ dependence under
rotation.
For continuous, uniform boresight coverage, its projection onto the $\cos2\psi$ and
$\sin2\psi$ columns vanishes unless $|m|=2$. In this limit, relative gain,
differential pointing, and differential beamwidth are suppressed by the
observing strategy itself, while differential ellipticity is not.

In practice the boresight coverage is not continuous. Each receiver observes at
4--8 boresight angles per season, so the suppression of the non-quadrupolar
modes is partial and depends on the distribution of observing time among boresight
angles. Finite boresight coverage therefore leaves residual leakage from
non-quadrupolar modes.

\subsection{Deprojection}
\label{sec:leakage:depj}

Deprojection is a filtering operation performed during mapmaking that removes
a selected high-leakage subspace of the differential beam
response.\cite{bicep2iii} We select this deliberately low-dimensional subspace
using beam measurements and instrumental understanding, and estimate leakage
from modes outside it with beam measurement-informed simulations.

The construction starts from a set of pair-difference beam modes
$K_i=\mathcal L_i B_0$, generated by linear operators $\mathcal L_i$ acting on
a beam model $B_0$. Applied to the beam-smoothed temperature sky
$\widetilde T\equiv T\ast B_0$, the same operators give the ``template maps''
$t_i=\mathcal L_i\widetilde T$, which are sampled along the real pointing,
then put through the same timestream processing and map binning as the data,
and used in a regression against the pair-difference
data.\cite{bicep2iii,bkvii} The fitted coefficients include
contributions from differential-beam leakage, polarized sky signal, noise, and
other systematics that project onto the templates; they do not in general equal
physical beam-mismatch amplitudes. Deprojection subtracts the fitted linear combination of the templates from
the timestream before the $Q$ and $U$ maps are formed, which removes the
selected subspace exactly. This construction describes timestream deprojection; in
beam map-space, we instead fit and subtract the selected beam modes $K_i$
directly from a measured pair-difference beam, leaving the ``undeprojected
residual'' used in Secs.~\ref{sec:results} and \ref{sec:basis}.

Neither formulation requires a particular beam parameterization. In the
mainline CMB analysis, $B_0$ is the azimuthally symmetrized,
receiver-averaged pair-sum beam, the real-space form of the receiver
$B_\ell$,\cite{bicep2iii,bkxi} and the $\mathcal L_i$ are the six
lowest-order spatial derivatives: the identity, the two first derivatives,
the Laplacian, and the two traceless second-derivative combinations. These
six have formed the deprojection basis of every mainline analysis since
BICEP2,\cite{bicep2iii} and we refer to them as the ``standard'' basis. Built
from the measured $B_\ell$, they retain the measured radial beam profile.

The familiar mode names in Table~\ref{tab:modes} follow from a special case.
Forming the standard modes from a Gaussian $B_0$ generates exactly the same
six-dimensional subspace as parametrizing the beam as an elliptical Gaussian
and perturbing its six parameters---gain, centroid, width, and
ellipticity---to first order. The derivatives of a Gaussian are Hermite
polynomials multiplying the same Gaussian, so the six templates span the
Hermite--Gauss modes on the two-dimensional beam plane through second order.
Regrouping them by azimuthal symmetry gives the equivalent Laguerre--Gauss
labeling used in Sec.~\ref{sec:basis:wide}. This correspondence gives the
conventional labels ``gain,'' ``pointing,'' ``beamwidth,'' and
``ellipticity,'' and the coefficients in the table. It is not required to
construct or apply the templates: for a non-Gaussian beam they remain well
defined, and only the equivalence to this particular parameter set breaks.

\begin{table}[tbp]
	\centering
	\caption{The standard six spatial-derivative templates and their
		elliptical-Gaussian correspondence. Here $\widetilde T$ is the temperature
		sky smoothed by the receiver $B_0$, $\sigma$ is the Gaussian width, and
		$p$ and $c$ are the plus and cross ellipticity components.}
	\label{tab:modes}
	\begin{tabular}{l l l l}
		\toprule
		\textbf{Mode}        & \textbf{Gaussian mismatch} & \textbf{Gaussian coefficient}  & \textbf{Template}                          \\
		\midrule
		Relative gain        & $g_A - g_B$                & $\delta g$                     & $\widetilde{T}$                            \\
		Pointing $x$         & $x_A - x_B$                & $\delta x$                     & $\nabla_x \widetilde{T}$                   \\
		Pointing $y$         & $y_A - y_B$                & $\delta y$                     & $\nabla_y \widetilde{T}$                   \\
		Beamwidth            & $\sigma_A - \sigma_B$      & $\sigma\,\delta\sigma$         & $(\nabla_x^2 + \nabla_y^2)\,\widetilde{T}$ \\
		Ellipticity $+$      & $p_A - p_B$                & $\frac{\sigma^2}{2}\,\delta p$ & $(\nabla_x^2 - \nabla_y^2)\,\widetilde{T}$ \\
		Ellipticity $\times$ & $c_A - c_B$                & $\frac{\sigma^2}{2}\,\delta c$ & $2\nabla_x \nabla_y\,\widetilde{T}$        \\
		\bottomrule
	\end{tabular}
\end{table}

\begin{figure}[tbp]
	\centering
	\includegraphics[width=0.55\linewidth, trim=0 0 0 60, clip]{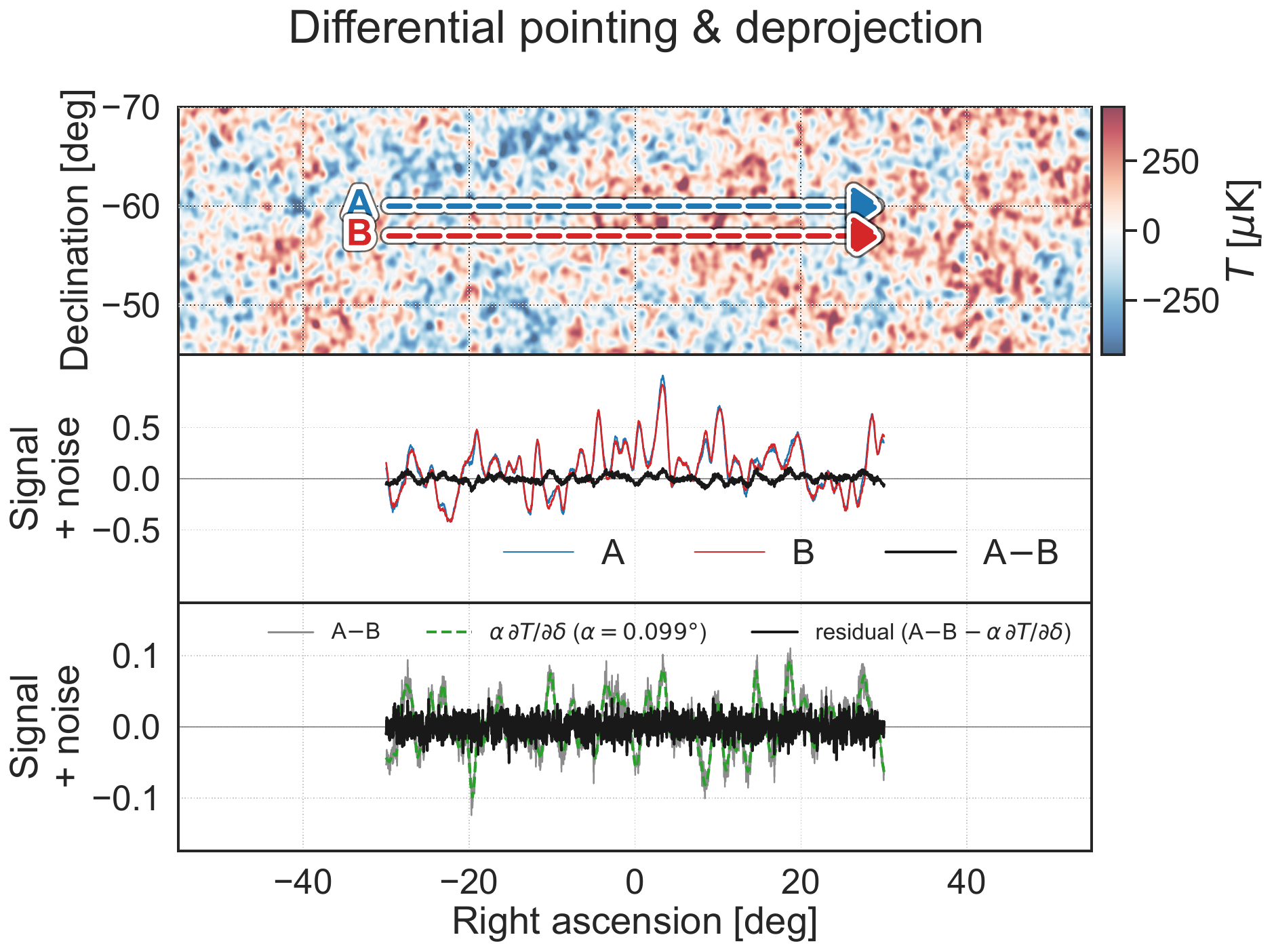}
	\caption{Pointing-driven \TtoP leakage and its removal. Two co-located
		detectors ($A,B$) with a small differential-pointing offset scan a simulated
		CMB temperature map (top), with noise injected into the $A,B$ timestreams.
		Their pair difference $A-B$ (middle) follows the gradient of the
		beam-smoothed temperature sky $\nabla\widetilde T$ along the offset
		direction, which is the \TtoP leakage for this mode. Regressing out the
		corresponding pointing template (bottom) leaves only noise, and the fitted
		coefficient recovers the injected offset.}
	\label{fig:cartoon}
\end{figure}

Each mode has a definite azimuthal order. Relative gain and beamwidth are
monopoles ($m=0$), the pointing modes are dipoles ($m=1$), and the ellipticity
modes are quadrupoles ($m=2$). These
symmetries set the boresight-rotation behavior of Sec.~\ref{sec:leakage:deck},
while the measured $B_\ell$ sets the radial structure. In
Fig.~\ref{fig:cartoon}, we illustrate deprojection for the simplest
nontrivial mode, differential pointing. A small centroid offset produces a
dipolar pair-difference beam, whose action on the temperature sky is the
directional derivative $\nabla\widetilde T$ along that offset. The regression
estimates its amplitude and subtracts the resulting pair-difference signal.

Mode selection balances leakage removal against filtering of the polarized
sky. Because the templates are built from the temperature sky and the true
polarization is correlated with it through $TE$, deprojection removes real
power along with the leakage, chiefly in $TE$ and $EE$ and chiefly for
differential ellipticity.\cite{bicep2iii,stgermaine2020} The mean filtering
is calibrated out by signal simulations that undergo the same
deprojection.\cite{bicep2iii} Through BK18 the ellipticity coefficients were
fixed from beam map measurements, so that only the mismatch contribution was
removed.\cite{bkxi} For BK24, we instead regress all six standard templates,
including differential beamwidth and differential ellipticity. For each
template, we fit one coefficient per detector pair and scan direction over one
observing phase.

Deprojection also carries a statistical cost. Removing the templates from the
timestreams reduces the amplitude of map modes that overlap them, raising the
sample variance and hence $\sigma(r)$. Preliminary simulations that repeat the
deprojection and E/B separation under each candidate configuration indicate
that the corresponding increase in $\sigma(r)$ is minor for $BB$, with further
work ongoing to assess this impact fully.
Structure outside the fitted subspace remains as the undeprojected residual,
which we estimate from the measured per-detector beams.

\section{FAR-FIELD BEAM MEASUREMENTS}
\label{sec:ffbm}

Characterizing the undeprojected residuals requires measurements of the
differential response of each detector pair. We therefore conduct dedicated
far-field beam mapping (FFBM) campaigns during most austral summers, measuring
the beams in situ through the same instrument configuration used
for CMB observations.

\subsection{Measurement configuration}
\label{sec:ffbm:setup}

\begin{figure}[tbp]
	\centering
	\captionsetup{skip=16pt}
	\resizebox{\linewidth}{!}{%
		\includegraphics[height=4cm]{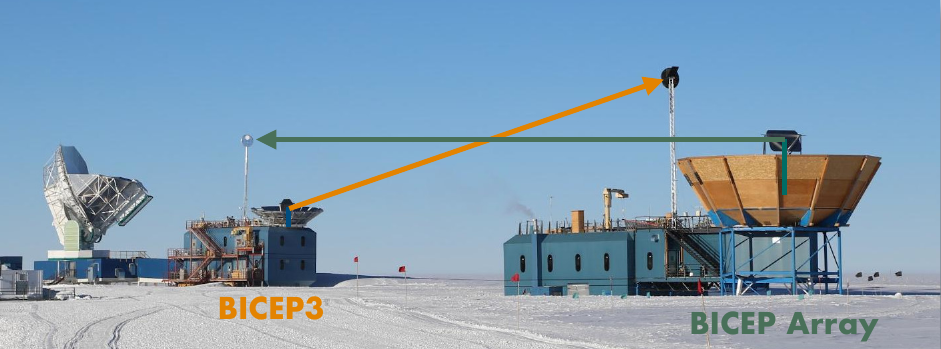}%
		\includegraphics[height=4cm]{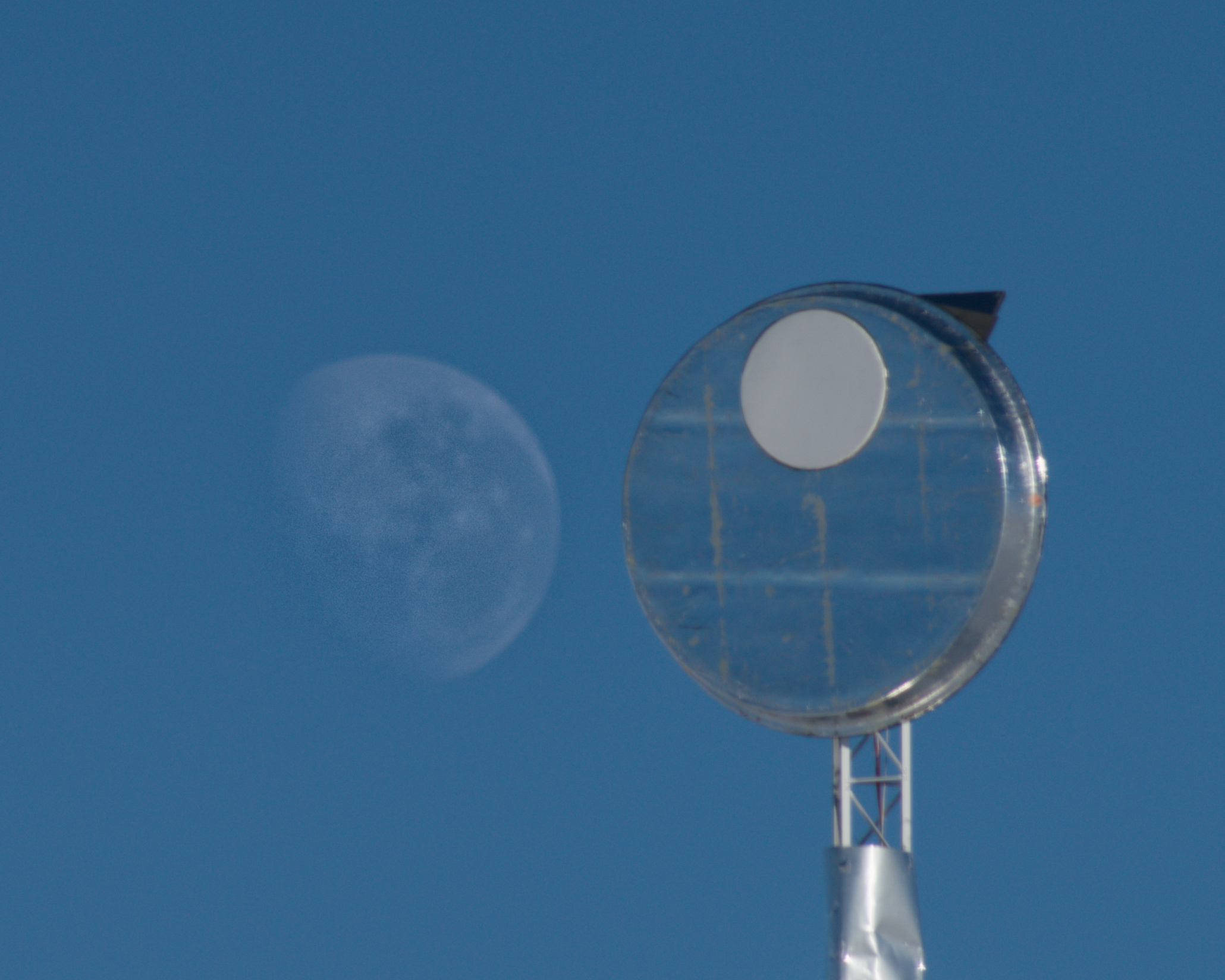}%
		\includegraphics[height=4cm]{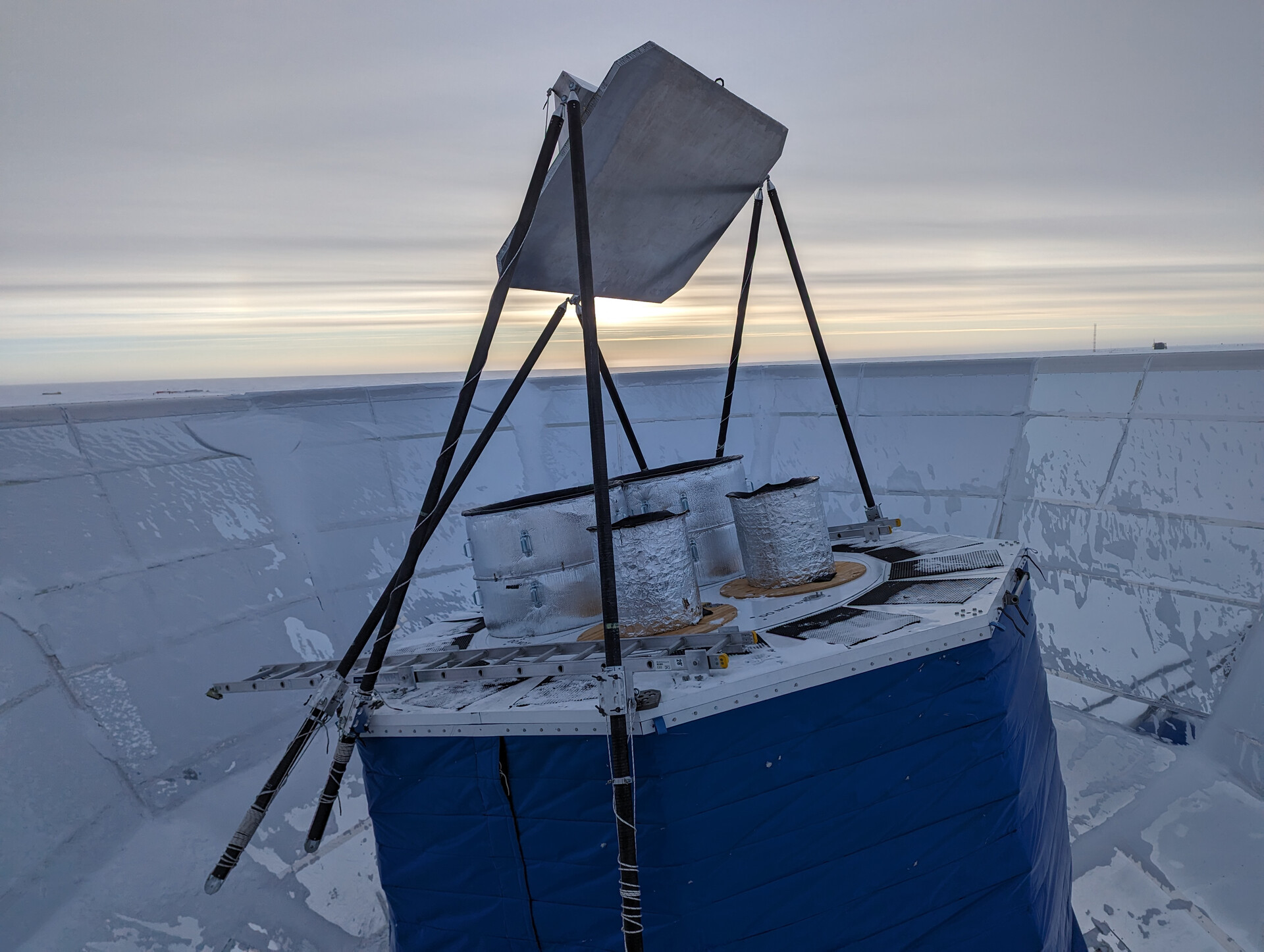}%
	}
	\caption{The far-field beam mapping configuration at Amundsen--Scott South
		Pole Station. Left: the Dark Sector Laboratory (DSL), housing BICEP3, and
		the Martin A. Pomerantz Observatory (MAPO), housing BICEP Array, separated
		by ${\sim}\qty{200}{\m}$; each telescope observes the source mounted on the
		mast of the opposite building (arrows), and the South Pole Telescope stands
		at far left. Center: the chopped thermal source on its mast. Right: the
		redirecting flat mirror mounted above the ground shield, which allows the
		receivers to see the source over the shield and near the horizon.}
	\label{fig:ffbmsetup}
\end{figure}

Beam mapping consists of observing a bright, unpolarized source in the far
field of the telescope and recording each detector's response as the
telescope scans across it. Our source is a chopped thermal emitter with a
\qty{61}{\cm} aperture, mounted on a mast approximately \qty{200}{\m} from
the telescope. The telescopes are designed to observe at elevations well
above the horizon during normal operations and are surrounded by a fixed
ground shield, so the receivers cannot see the source directly. A large
flat redirecting mirror
(the ``far-field flat,'' FFF) is mounted above the ground shield at
approximately \qty{45}{\degree} so that the boresight, pointed near zenith, is
folded toward the horizon and onto the source (Fig.~\ref{fig:ffbmsetup}).

The source is chopped at ${\sim}\qty{16}{\Hz}$ between an ambient-temperature
load and cold zenith sky, viewed through a redirecting mirror behind the source
aperture. The telescope rasters in azimuth and elevation across the source
while the detector timestreams are demodulated at the chop frequency. The
demodulation confines the measurement to a narrow band around the chop
frequency, above the low-frequency noise, reducing sensitivity to $1/f$
fluctuations and slow drifts. A single raster yields one
``component'' beam map per detector. We repeat the raster over several
boresight angles and mirror positions, so that a given detector is
measured with its beam landing on different parts of the FFF and at different
orientations relative to the instrument. A full campaign produces of order
\num{20000} per-detector maps across all receivers and bands. The boresight
coverage helps us distinguish structure fixed in the instrument from
structure fixed in the measurement configuration. It also samples the
rotation over which the CMB observations are coadded.

\subsection{Beam products}
\label{sec:ffbm:products}

We fit a two-dimensional elliptical Gaussian to the timestreams of each
component before binning. The fitted center, width, and ellipticity summarize
the beam at low order. They provide the beam center used when combining map measurements, the
reference for the automatic beam-measurement cuts, and the measured per-pair
differential parameters of the Gaussian correspondence in
Table~\ref{tab:modes}. To ensure a baseline beam quality, we apply geometric cuts, fit-quality cuts, and shape cuts referenced to the
Gaussian fit. The geometric cuts reject measurements for which the beam
footprint is not fully contained on the FFF.

To form our best estimate of each detector's beam response, we recenter the
surviving component maps on their fitted beam centers, normalize them to peak,
and take the median across measurements. We call the result the detector's
``composite'' beam map. The median reduces sensitivity to transient pickup,
source-mast reflections, and passing structures.

Coadding the composite beam maps across detectors gives the array-averaged beam from
which the beam transfer function $B_\ell$ is derived.
A \qty{61}{\cm} aperture at \qty{200}{\m} subtends
approximately \qty{10.5}{\arcminute}, a nontrivial angular scale compared
with the receiver beams. The measured maps therefore include convolution with
the source disk. We deconvolve its window function from the $B_\ell$ in Fourier
space. Because the source subtends a fixed angle while
the beams narrow with frequency, this correction becomes progressively more
important toward the higher bands.

\section{BEAM MEASUREMENT-INFORMED SIMULATIONS}
\label{sec:beamsims}

To estimate the residual \TtoP leakage in the CMB data, we use the measured
per-detector composite beam maps as inputs to a specialized set of timestream
simulations, which we call beam measurement-informed simulations. Each
composite beam is
convolved with a
temperature-only \textit{Planck} sky and sampled along the real pointing,
reproducing the scan strategy, boresight-angle distribution, and per-detector
observing time. The simulated timestreams then undergo the same filtering,
cuts, deprojection, and mapmaking as the data, producing simulated $Q$ and $U$
maps.

Each simulation is specified by a beam-mapping measurement epoch $b$ and a CMB
observing year $y$. An epoch groups the beam mapping campaigns of
Sec.~\ref{sec:ffbm} over which the receiver optics were unchanged, and is
distinct from the CMB observing year to which the resulting beams are applied.
Within the simulation, the detector composite beams from epoch $b$ are
differenced by pair. We denote the resulting collection of
per-pair difference beams by $\Delta B_b$. The year-$y$ observation and
reduction operator $\mathcal{P}_y$ includes the scan strategy, coverage,
timestream filtering, channel cuts, deprojection, and mapmaking. Acting on
$\Delta B_b\ast T$, it returns the two-component Stokes leakage map
$L_{b,y}\equiv(Q^{\rm leak}_{b,y},U^{\rm leak}_{b,y})$:
\begin{equation}
	L_{b,y} = \mathcal{P}_y\!\left[\, \Delta B_b \ast T \,\right] .
	\label{eq:beamsim-map}
\end{equation}
For a given receiver, the temperature field $T$ is a fixed, beam-deconvolved
\textit{Planck} temperature map at the nearest frequency, restricted to the
observing field. The input sky has $Q=U=0$ and no injected noise; we use the
same temperature map for every year and epoch. The simulated timestreams
undergo the same filtering and deprojection as the real data, with deprojection
templates built from the same external temperature map. At fixed observing
year, variation between epochs comes from beam-measurement noise and changes in
the measured $\Delta B_b$. Since the input sky has no polarization, any output
$Q$ or $U$ is residual beam leakage.

For a compact figure of merit, we fit the residual $BB$ spectrum to the shape
of a primordial tensor signal. We collect the residual bandpowers for a single
receiver band into the vector $\hat{\mathbf{c}}$, after correcting for
filtering and beam suppression.
For the values quoted here, $\mathbf{m}$ contains the expected bandpowers of
the theoretical tensor $BB$ spectrum at a fiducial $r_{\rm fid}$, projected
through the corresponding bandpower window functions. We take $\mathbf{N}$ to
be diagonal, with entries given by the noise-only bandpower variances of the
corresponding real analysis, and define
\begin{equation}
	\rho \;=\; r_{\rm fid}\;
	\frac{\mathbf{m}^{\sf T}\mathbf{N}^{-1}\hat{\mathbf{c}}}
	{\mathbf{m}^{\sf T}\mathbf{N}^{-1}\mathbf{m}} .
	\label{eq:rho}
\end{equation}
Equation~\ref{eq:rho} is the inverse-variance-weighted least-squares
amplitude obtained by fitting the residual spectrum to the tensor spectrum.
The prefactor expresses this amplitude in units of $r$.

The quantity $\rho$ is a single-frequency-band spectral figure of merit and is
independent of the cosmological likelihood. The bias $\Dr$, by contrast, is
the shift in
the recovered tensor-to-scalar ratio after full multi-frequency weighting and
component separation. The contribution of a given leakage spectrum to $\Dr$
depends on the survey weight of that receiver. A receiver with
$\rho>0.001$ can therefore remain consistent with the
$\Dr\lesssim0.001$ requirement if its survey weight is low. The systematic
requirement must be tested through likelihood propagation.

Two properties of the beam measurements motivate the epoch-cross estimator
used in Sec.~\ref{sec:results}. The first is beam map measurement noise. The
individual
component maps have noise floors between roughly $-18$ and $-35$\,dB
relative to peak, depending strongly on the detector and the observing
conditions, and some of that noise persists in each composite beam map. Because the
leakage power is quadratic in the pair-difference beam, this residual noise
produces a positive bias in an auto-spectrum rather than averaging away. The
second is temporal coverage. The available beam measurements do not cover
every observing year, so assigning a measured beam to an
unmeasured season assumes stability over that interval. In cross-spectra
between independent measurement epochs, uncorrelated measurement noise has
zero mean. The scatter among these cross-spectra provides a test of beam
stability over intervals without measurements.

\section{RESIDUAL BEAM MISMATCH AND \texorpdfstring{\TtoP}{T-to-P} LEAKAGE FOR BK24}
\label{sec:results}

\subsection{The residual in beam map-space}
\label{sec:results:beamspace}

\begin{figure}[tbp]
	\centering
	\begin{subfigure}[t]{0.49\linewidth}
		\centering
		\includegraphics[width=\linewidth]{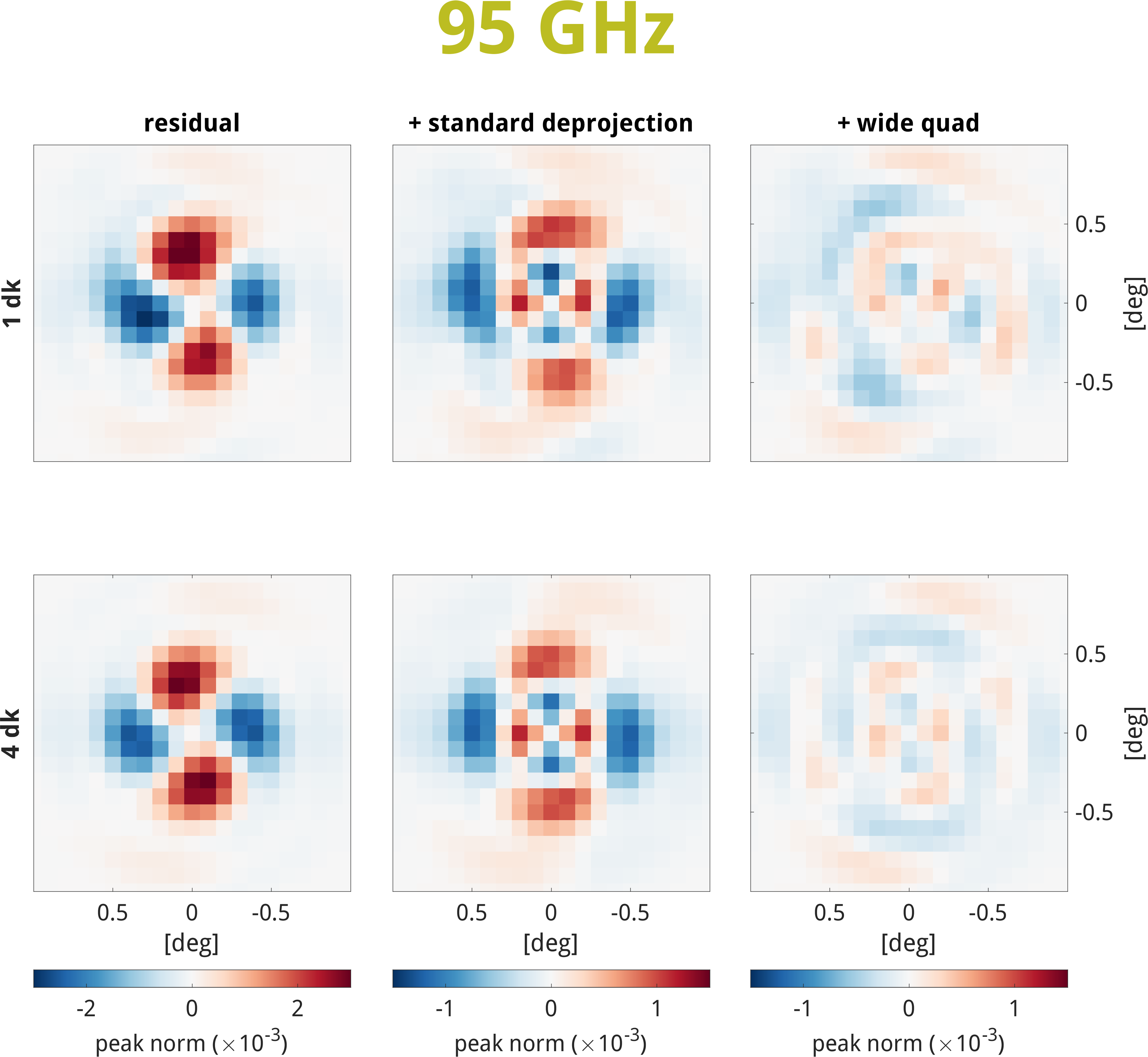}
		\caption{BICEP3, 2018, coadded over four boresight angles.}
		\label{fig:dkcompare:b3}
	\end{subfigure}
	\hfill
	\begin{subfigure}[t]{0.49\linewidth}
		\centering
		\includegraphics[width=\linewidth]{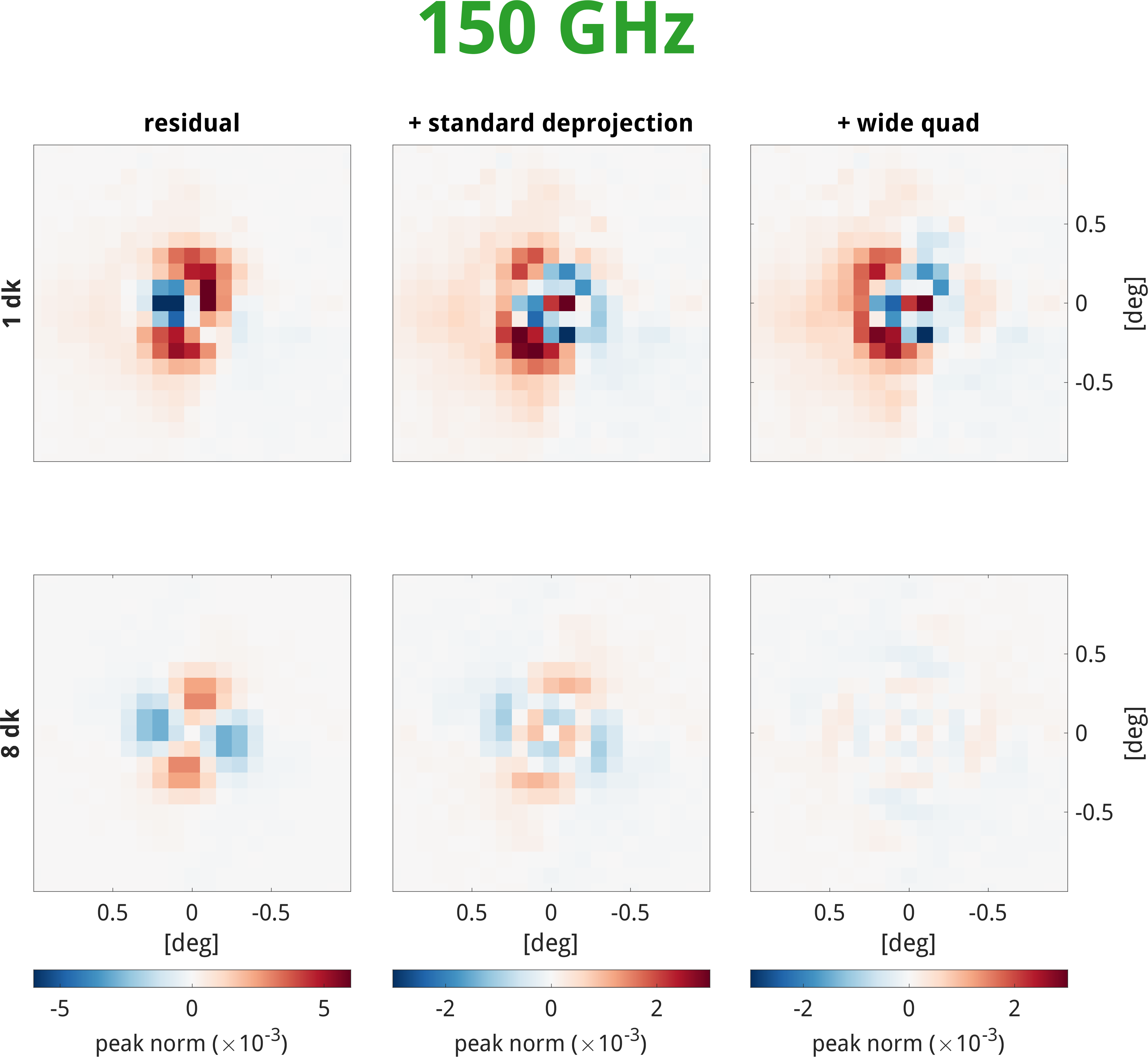}
		\caption{BA2-150, 2024, coadded over eight boresight angles.}
		\label{fig:dkcompare:ba2}
	\end{subfigure}
	\captionsetup{skip=16pt}
	\caption{Array-averaged pair-difference beams, peak-normalized to the pair-sum
		beam, for the BICEP3 \qty{95}{\GHz} receiver (\subref{fig:dkcompare:b3}) and
		the BICEP Array \qty{150}{\GHz} receiver (\subref{fig:dkcompare:ba2}). Within
		each panel the columns show the raw pair-difference beam, the beam after
		standard six-mode deprojection, and the beam after additionally
		deprojecting the wide quadrupole. The rows compare a single boresight angle
		against the full boresight coadd. Coadding over boresight angle suppresses the non-quadrupolar
		structure visible in the single-angle maps. Standard deprojection removes the
		compact quadrupole but leaves a wide quadrupole at the diffraction scale,
		which the wide-quadrupole template removes.}
	\label{fig:dkcompare}
\end{figure}

Figure~\ref{fig:dkcompare} shows how boresight rotation and deprojection change the
array-averaged pair-difference beams of the BICEP3 and BA2-150 receivers. The comparison separates structure suppressed by
the observing strategy from structure removed by deprojection.

The raw pair-difference beam is dominated by a compact quadrupole. As expected
from the symmetry argument of Sec.~\ref{sec:leakage:deck}, boresight coaddition
reduces the visible monopolar and dipolar structure but not the quadrupolar
response. Standard deprojection removes the compact quadrupole, leaving
coherent quadrupolar structure at larger angular scales. This residual has a radial profile that
the compact second-derivative templates of Table~\ref{tab:modes} cannot
reproduce at any amplitude, and its angular scale is comparable to the
diffraction structure of the aperture.
Section~\ref{sec:basis} constructs the corresponding wide-mode templates
and presents their instrumental interpretation.

\subsection{\texorpdfstring{BB leakage spectra}
	{Leakage spectra and the equivalent-r statistic}}
\label{sec:results:spectra}

Beam-measurement noise remaining in a composite beam map biases its leakage
auto-spectrum high. We therefore use cross-spectra between independent
measurement epochs, for which uncorrelated measurement noise has zero
mean. We divide the BICEP3 beam
measurements into four high-signal-to-noise beam-mapping epochs---2016, the
combined 2017/2018 campaigns, 2019, and 2023. For each epoch
$b$, we form the set $\{L_{b,y}\}$ over the observing years
$y=2016$--$2024$ and coadd it into a single nine-year leakage map $L_b$. The
four $L_b$ estimate the same leakage to the extent that the differential beam
is stable between epochs. Averaging the six pairwise $BB$ cross-spectra
estimates the leakage common to the epochs. We estimate the uncertainty with a
delete-one-epoch jackknife, a resampling estimate in which we recompute the
mean with each epoch left out in turn and take the spread of the four results.
This accounts for the correlation among the
six pairs, since each epoch appears in three of them, and provides a measure
of the consistency of the residual across epochs.

\begin{figure}[tbp]
	\centering
	\includegraphics[width=0.62\linewidth]{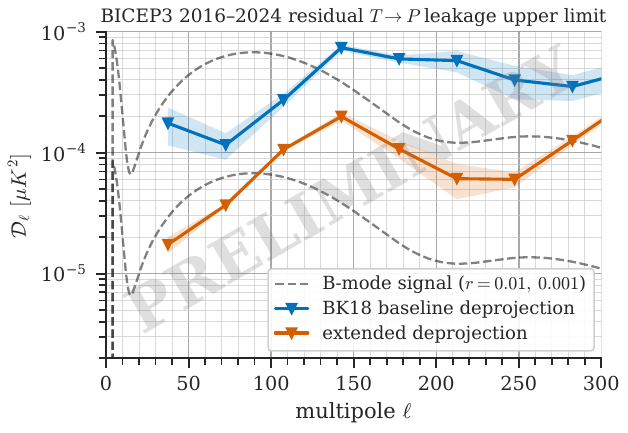}
	\caption{Residual \TtoB leakage $\mathcal{D}_\ell^{BB}$ from beam
		measurement-informed simulations of BICEP3, 2016--2024, under the BK18 baseline deprojection
		(blue) versus the extended deprojection basis, which deprojects all six
		standard templates, their radially smoothed counterparts, and
		readout-crosstalk templates (orange).
		Each curve is the mean of the six pairwise cross-spectra between the four
		measurement epochs; beam-measurement noise that is uncorrelated between
		epochs has zero mean in this estimator. The shaded bands show the
		delete-one-epoch jackknife uncertainty. The dashed curves are the tensor $BB$
		spectrum for $r = 0.01$ and $r = 0.001$, given as reference. The curves are labeled as
		preliminary upper limits because measurement-side contributions can raise the
		simulated leakage above that realized in the data.}
	\label{fig:bbresid}
\end{figure}

Figure~\ref{fig:bbresid} compares the epoch-cross $BB$ spectra for the
BK18 baseline deprojection (Sec.~\ref{sec:leakage:depj}) and the extended
deprojection basis. The latter deprojects all six standard templates, their
radially smoothed counterparts, and readout-crosstalk templates
(Sec.~\ref{sec:basis}). Table~\ref{tab:rho} reports $\rho$ for these
cross-spectra and the corresponding mean single-epoch auto-spectra. Extending the deprojection basis reduces the residual leakage by
\qtylist{75;77}{\percent} in the cross- and auto-spectra, respectively. The cross-spectrum values are
\qtylist{83;92}{\percent} of
the corresponding auto-spectrum values, indicating that most of the simulated leakage
is reproduced across independent epochs rather than set by the noise of one
epoch.

\begin{table}[tbp]
	\centering
	\caption{Residual \TtoB leakage for BICEP3, 2016--2024, expressed through
		Eq.~\ref{eq:rho} as an equivalent tensor-to-scalar ratio. The auto column is
		the mean of the four single-epoch auto-spectra and includes the
		beam-measurement noise. The cross column is the mean of the six pairwise
		epoch cross-spectra, in which measurement noise uncorrelated between
		epochs has zero mean. The reported cross-spectrum uncertainty is
		estimated with the delete-one-epoch jackknife.}
	\label{tab:rho}
	\begin{tabular}{l c c}
		\toprule
		\textbf{Deprojection} & \textbf{Auto [$\times 10^{-3}$]} & \textbf{Cross [$\times 10^{-3}$]} \\
		\midrule
		BK18 baseline         & 5.4                              & $4.5 \pm 0.7$                     \\
		Extended basis        & 1.22                             & $1.12 \pm 0.06$                   \\
		\bottomrule
	\end{tabular}
\end{table}

These single-band $\rho$ values are not the
likelihood bias $\Dr$ to which the BK24 requirement directly applies. The
propagation of single-band leakage into $\Dr$ through
multi-frequency weighting and component separation is not evaluated here.
The simulations also contain the measurement-side contributions of
Sec.~\ref{sec:ffbm}, which are not present in the on-sky beam. Direct
validation requires cross-spectra between the simulated leakage maps and the
real data, which need unblinded maps and remain pending. In earlier datasets,
these cross-spectra were consistent in shape with the simulation auto-spectra
but lower in amplitude. In the BK18 analysis, the cross-spectrum amplitude was
roughly half the simulation auto-spectrum amplitude, although the predicted
leakage was not conclusively detected.\cite{bkxi,stgermaine2021} Each of these
considerations places the expected on-sky bias below the quoted $\rho$, so the
extended basis puts BICEP3 on target to meet the BK24 requirement on
differential beam response.

The agreement among epochs is consistent with the receiver optical chain
setting the differential beams. Together with the
per-band multicomponent analysis of Vergès et al.,\cite{verges2026} this
result supports prioritizing BICEP3 for additional beam-leakage mitigation in
BK24.

\section{EXTENDING THE DEPROJECTION BASIS}
\label{sec:basis}

The increased sensitivity of the BK24 polarized maps motivates us to identify,
model, and remove residual structure outside the standard deprojection basis.
We introduce two additional template families to remove the leading
residuals. The wide modes add radial structure at each azimuthal order
represented by the standard templates, while the second family describes
readout crosstalk. A forward optical model relates the wide modes to
first-order perturbations of instrument parameters.
Table~\ref{tab:depj} summarizes these extensions.

\begin{table}[tbp]
	\centering
	\caption{Additional templates in the extended deprojection basis used for the
		BK24 analysis. Here $\widetilde{T}_{\rm w}$ denotes the temperature map
		after the common axisymmetric radial smoothing used for all six wide-mode
		templates. The crosstalk template is the multiplexing partner's
		beam-smoothed temperature map evaluated along the partner's pointing.}
	\label{tab:depj}
	\begin{tabular}{l l}
		\toprule
		\textbf{Mode}           & \textbf{Templates}                                                                             \\
		\midrule
		Wide monopole ($m=0$)   & $\widetilde{T}_{\rm w},\ (\nabla_x^2 + \nabla_y^2)\,\widetilde{T}_{\rm w}$                     \\
		Wide dipole ($m=1$)     & $\nabla_x \widetilde{T}_{\rm w},\ \nabla_y \widetilde{T}_{\rm w}$                              \\
		Wide quadrupole ($m=2$) & $(\nabla_x^2 - \nabla_y^2)\,\widetilde{T}_{\rm w},\ 2\nabla_x \nabla_y\,\widetilde{T}_{\rm w}$ \\
		Readout crosstalk       & multiplexing partner's $T$                                                                     \\
		\bottomrule
	\end{tabular}
\end{table}

\subsection{The wide modes}
\label{sec:basis:wide}

Truncating a Gaussian illumination profile at an aperture gives the beam two
characteristic optical scales, one set by the width of the Gaussian and the
other by the size of the aperture. In the far field,
these scales appear primarily in the compact main beam and diffraction-scale
structure, respectively. For each standard differential operator, the measured pair-sum beam
$B_\ell$ fixes a single radial profile containing both scales, so one fitted
amplitude cannot vary their contributions independently. The residual of
Sec.~\ref{sec:results:beamspace} has the azimuthal order of the compact
quadrupole but a broader radial profile, which the standard ellipticity
templates cannot reproduce at any amplitude. We introduce a second radial profile through a
broader axisymmetric smoothing of the temperature map. Applying the same six
differential operators to the more broadly smoothed map preserves the
azimuthal order of each operator and gives the six wide-mode templates of
Table~\ref{tab:depj}.

At fixed azimuthal order $m$, successive radial profiles can be orthogonalized
against the lower-order profiles to define a radial order $j$. For a Gaussian
fiducial, this construction gives the Laguerre--Gauss family. In this labeling
the standard six occupy $(j,m)=(0,0)$ and $(1,0)$ for relative gain and
differential beamwidth, $(0,1)$ for the two pointing modes, and $(0,2)$ for
the two ellipticity modes; the wide modes add $(1,1)$, $(1,2)$, and $(2,0)$.
The two wide monopole templates together add a single radial mode, since the
standard pair already spans $j=0$ and $j=1$ at $m=0$. Thus the wide templates
extend the deprojection basis with an additional radial profile at each
azimuthal order represented by the standard templates.

For BICEP3, we define the wide modes using a Gaussian smoothing of FWHM
\qty{31.2}{\arcminute}, which was chosen to roughly match the measured quadrupolar residual. These
six templates form part of the extended deprojection basis used for the BK24
analysis. The physical-optics model in the following section relates this
empirical construction to perturbations of the pupil illumination, phase, and
aperture. It could instead define the wide modes directly, setting their radial
scales from the modeled optics without fitting an empirical smoothing width
and providing a common prescription across receivers.

\subsection{A forward optical model}
\label{sec:basis:model}

To relate these additional radial profiles to instrument parameters, we
differentiate a forward optical model of a truncated, Gaussian-illuminated
aperture. We describe its generalized pupil
function\cite{goodman2017}
\begin{equation}
	P(\mathbf{r}; \nu, \boldsymbol{\theta})
	= \underbrace{g\,A(\mathbf{r}; \sigma, e_+, e_\times, \nu)}_{\text{Gaussian taper}}
	\;\cdot\;
	\underbrace{\Pi(\mathbf{r}; r_{\rm ap}, e^{\rm ap}_+, e^{\rm ap}_\times)}_{\text{aperture truncation}}
	\;\cdot\;
	\underbrace{e^{-i\frac{2\pi\nu}{c}\left(\boldsymbol{\delta}\cdot\mathbf{r} + W(\mathbf{r};\mathbf{Z})\right)}
		\vphantom{e_\times}}_{\text{pointing and phase aberration}} ,
	\label{eq:pupil}
\end{equation}
propagated to the far field as
\begin{equation}
	B(\hat{n}_0; \boldsymbol{\theta}) =
	\underbrace{\int\! d\nu\, \bar{w}(\nu - \nu_0)}_{\text{bandpass}}
	\;\underbrace{S(\nu)\!\int\! d^2\hat{n}\, I_s(\hat{n})}_{\text{source}}
	\;\underbrace{\left| \iint\! d^2\mathbf{r}\; P\, e^{-i\frac{2\pi\nu}{c}\,\mathbf{r}\cdot(\hat{n}-\hat{n}_0)} \right|^2}_{\text{Fraunhofer}} .
	\label{eq:farfield}
\end{equation}
Here $\mathbf{r}$ is a two-dimensional pupil coordinate and $P$ is the
complex pupil field. The illumination amplitude $A$ is truncated by the
aperture support $\Pi$. Pointing $\boldsymbol{\delta}$ and the Zernike
coefficients $\mathbf{Z}$ enter through the linear and aberrated wavefront
terms. Together with the field-amplitude normalization $g$, illumination
width and ellipticities $(\sigma,e_\pm)$, and aperture radius and
ellipticities $(r_{\rm ap},e^{\rm ap}_\pm)$, these quantities form the
parameter vector $\boldsymbol{\theta}$. The Fraunhofer term gives the
monochromatic intensity response at $\hat n_0$ to source direction $\hat n$.
The measured instrument response $\bar w$ is averaged separately from the
source spectrum $S(\nu)$, while $I_s$ describes the angular profile of the
chopped source. This construction matches the measurement configuration of
Sec.~\ref{sec:ffbm}. The finite source distance enters the model as a
quadratic (defocus) phase in $W$, at most a tenth of a wave at the aperture
edge, with subpercent effects on the modeled FWHM for both BICEP3 and
BA2-150.

\begin{figure}[tbp]
	\centering
	\includegraphics[width=0.78\linewidth]{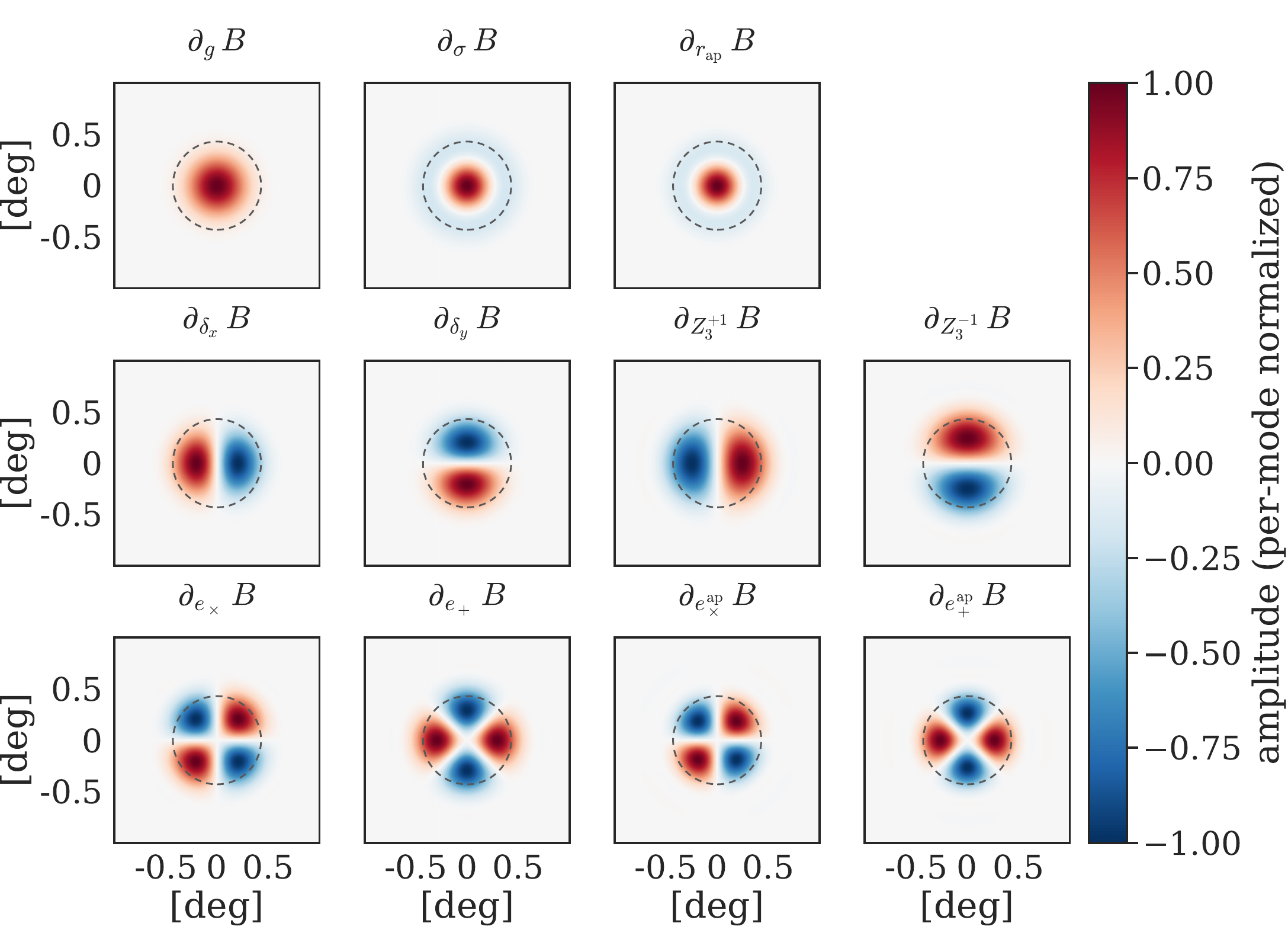}
	\caption{First-order far-field beam responses obtained by differentiating
		Eq.~\ref{eq:farfield} with respect to the modeled physical parameters and
		evaluating the derivatives at the fiducial parameter set $\boldsymbol{\theta}_0$
		for the \qty{95}{\GHz} beam. Each panel is a normalized two-dimensional beam map and one
		column of the Jacobian $\left.\partial B/\partial\boldsymbol{\theta}\right|_{\boldsymbol{\theta}_0}$. The model is implemented in
		dLux,\cite{desdoigts2023} which supports automatic differentiation through
		JAX.\cite{jax2018} The response maps are arranged by azimuthal order
		(monopole, dipole, quadrupole from top to bottom). The first two columns
		show the responses to gain, pointing, illumination width, and
		illumination ellipticity, the six physical perturbations conventionally
		associated with the standard template subspace. The remaining columns
		show the responses to the aperture radius, the $Z_3^{\pm1}$ pupil-phase
		components (the third-order, spin-1 Zernike polynomials, in classical
		optics the coma aberration), and aperture
		ellipticity, which are not fully contained in that subspace. The dashed
		circle marks the first Airy null of the aperture,
		\qty{0.43}{\degree} at band center, separating main-beam scales from
		diffraction scales.}
	\label{fig:modegrid}
\end{figure}

For small offsets between the two detectors' parameters, the
pair-difference beam is a first-order perturbation about a fiducial
parameter set:
\begin{equation}
	B(\boldsymbol{\theta}_0 + \delta\boldsymbol{\theta})
	- B(\boldsymbol{\theta}_0)
	\simeq
	\sum_i
	\left.\frac{\partial B}{\partial\theta_i}\right|_{\boldsymbol{\theta}_0}
	\delta\theta_i .
	\label{eq:firstorderbeam}
\end{equation}
The columns of this Jacobian describe the first-order beam response to the
modeled instrument parameters. We use these response maps as candidate
modes when evaluating or extending the deprojection subspace
(Fig.~\ref{fig:modegrid}).

\begin{figure}[tbp]
	\centering
	\includegraphics[width=1.0\linewidth]{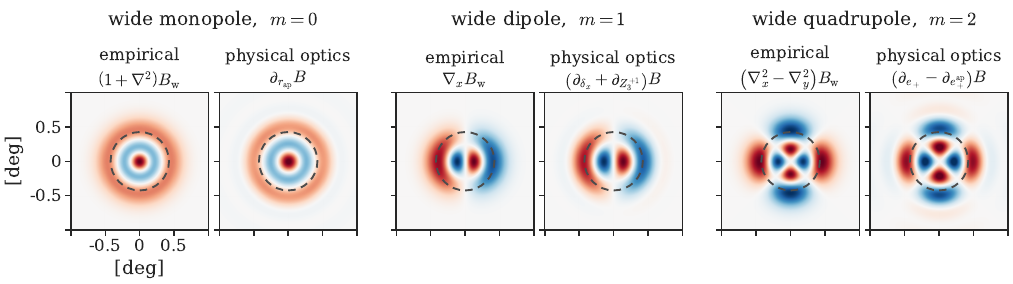}
	\captionsetup{skip=10pt}
	\caption{Empirical (left) and physical-optics (right) constructions of
		the wide monopole ($m=0$), dipole ($m=1$), and quadrupole ($m=2$); one
		orientation is shown for each $m\ne0$ mode. The empirical modes apply the
		standard differential operators to a Gaussian beam $B_{\rm w}$ with FWHM
		\qty{31.2}{\arcminute}, a width chosen to roughly match the residual quadrupole. The physical-optics modes are formed from normalized
		derivatives with respect to the instrument parameters labeled above each
		panel. The standard six-mode subspace is projected out in every panel. The dashed circle marks
		the first Airy null, \qty{0.43}{\degree} at band center.}
	\label{fig:widemodes}
\end{figure}

The modeled responses separate into the azimuthal blocks shown in
Fig.~\ref{fig:modegrid}, allowing the physical-optics and standard deprojection
subspaces to be compared independently at each $m$. After the standard six
modes are projected out, linear combinations of the dipole and quadrupole
responses cancel their compact components and leave the additional radial
structure shown in Fig.~\ref{fig:widemodes}.

The radial ordering makes the wide-dipole identification especially direct:
removing differential pointing from the pupil-phase coma response gives the
next dipole radial order. At $m=0$, aperture truncation introduces an
additional diffraction-scale response beyond gain and beamwidth;
Fig.~\ref{fig:widemodes} uses the residual aperture-radius response.

\subsection{Near-field comparison}
\label{sec:basis:nearfield}

The physical-optics construction associates the wide dipole with offset
aperture illumination. We test this interpretation using near-field beam maps,
which provide an independent measurement of the illumination offset. For each
BICEP3 detector pair, let
$\boldsymbol{\mu}_A$ and
$\boldsymbol{\mu}_B$ be the measured beam-centroid vectors in the near-field
source plane. We define the differential near-field beam steer as
\begin{equation}
	\Delta\boldsymbol{\mu}
	\equiv \boldsymbol{\mu}_A-\boldsymbol{\mu}_B
	\in\mathbb{R}^2 .
	\label{eq:diffnfbmsteer}
\end{equation}
Common displacements cancel in this difference, leaving a quantity which measure differences in aperture illumination. After projecting out the standard modes in the
far field, we fit the two wide-dipole orientations to each residual
pair-difference beam and assign the coefficient vector
$\boldsymbol{c}=(c_x,c_y)$.

\begin{figure}[tbp]
	\centering
	\includegraphics[width=0.70\linewidth]{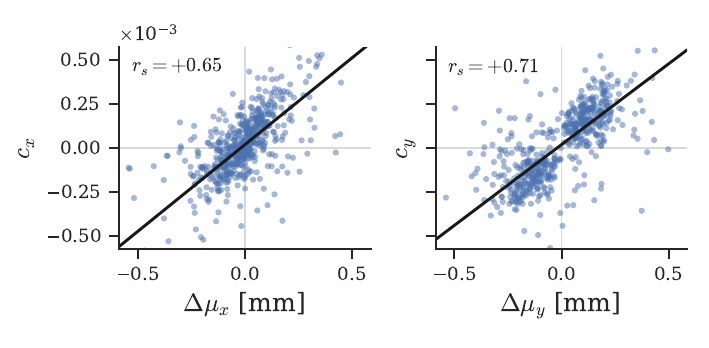}
	\caption{Differential near-field beam steer compared with the fitted
		far-field wide-dipole coefficient. Each
		point represents one BICEP3 detector pair, the black lines are the
		median of the pairwise slopes, and $r_s$ is the Spearman rank
		correlation. The two
		components of
		$\Delta\boldsymbol{\mu}$ have been rotated into the far-field coefficient
		convention for the component-wise comparison.}
	\label{fig:coma-steer}
\end{figure}

The differential near-field beam steer is correlated with the far-field
coefficient vector (Fig.~\ref{fig:coma-steer}). For the 559 detector pairs
for which the near-field beam data passes quality cuts, the mean of the two component-wise Spearman
correlations is $r_s=0.68$. The correlation supports an association between
the wide dipole and differences in offset aperture illumination within
detector pairs.

\subsection{Readout crosstalk}
\label{sec:basis:xtalk}

The \BK detectors are read out with time-division SQUID multiplexing, in which
detectors sharing a multiplexing column are addressed in sequence by row. Finite
isolation between rows means that a detector's timestream contains scaled
copies of the signals of the rows addressed immediately before and after
it, its upstream and downstream multiplexing partners. Since a partner is a
different detector at a different focal-plane position, its beam points
elsewhere on the sky, and the crosstalk appears in the pair-difference beam
as a faint displaced copy of the partner's beam. We construct the
corresponding templates from the partners' beam-smoothed temperature maps,
sampled along each partner's pointing.

\begin{figure}[tbp]
	\centering
	\includegraphics[width=0.46\linewidth]{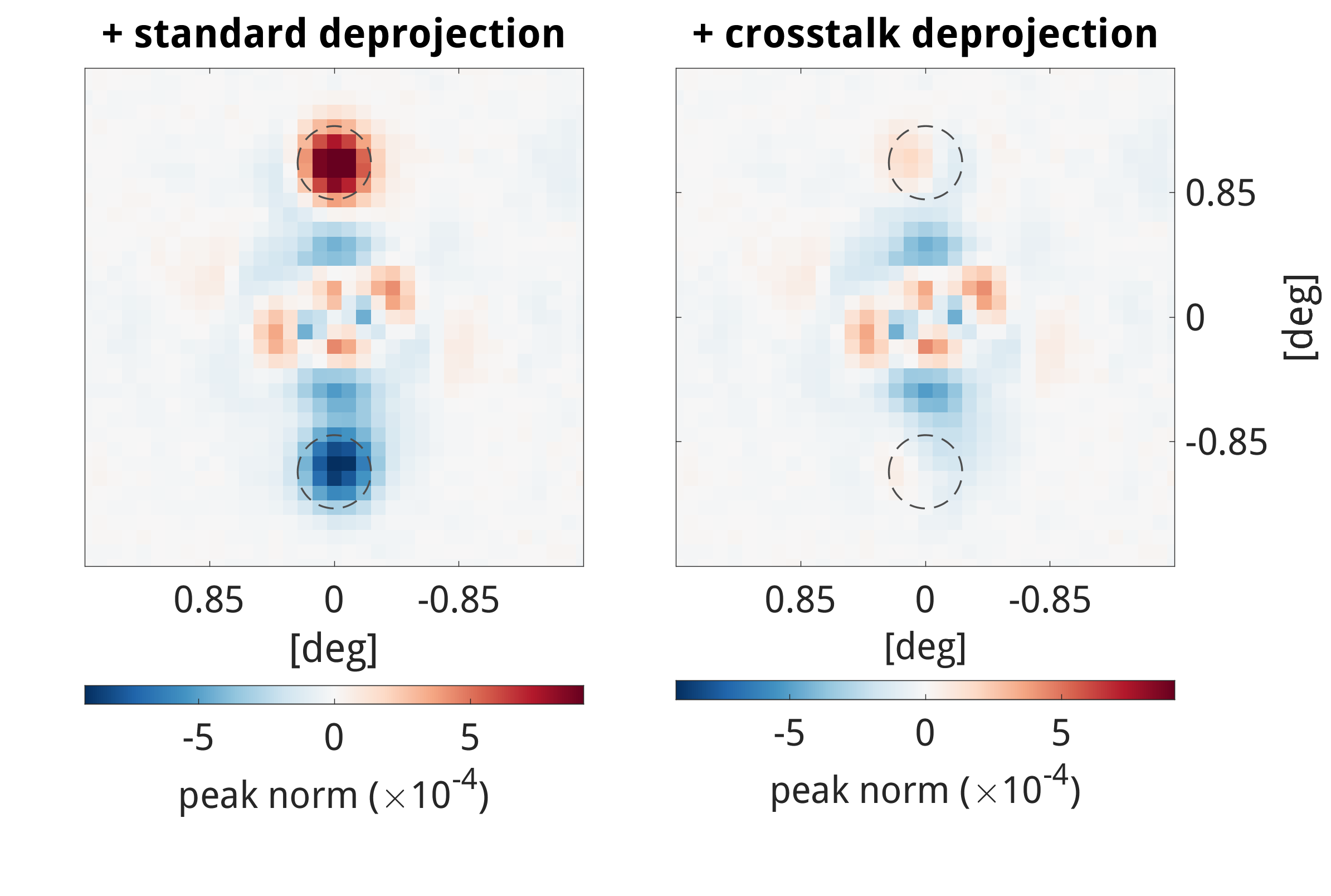}
	\captionsetup{skip=10pt}
	\caption{Readout crosstalk in BICEP3 at \qty{95}{\GHz}. Residual
		pair-difference beams, coadded so that the crosstalk partners of every
		pair align, after standard deprojection (left) and after additionally
		deprojecting the crosstalk templates (right). The coadd is illustrative,
		constructed to accentuate the crosstalk rather than to represent any
		detector population. Dashed circles mark the modeled partner locations,
		where the crosstalk partner beams appear in the left panel and are
		substantially reduced in the right.}
	\label{fig:xtalk}
\end{figure}

Positive and negative crosstalk partner beams appear at the upstream and
downstream locations at a few parts in $10^4$ of the main beam
(Fig.~\ref{fig:xtalk}). Adding the crosstalk templates substantially reduces
these features with little change to the surrounding residual. Among pairs
with a significant detection, the coefficients cluster at \qty{0.11}{\percent} of the
pair-sum beam peak, with a scatter of \qty{0.05}{\percent}, and are consistent
across the four readout crates.

\section{CONCLUSIONS AND OUTLOOK}
\label{sec:conclusion}

Differential beam response is potentially a leading source of instrumental systematic error
in $r$ for the \BK program. We estimate the residual \TtoP leakage with a
specialized set of end-to-end timestream simulations, which we call beam
measurement-informed simulations. These simulations propagate in situ, per-detector far-field beam
measurements through the BK analysis pipeline.

For BICEP3 over 2016--2024, the epoch-cross spectra give a preliminary
upper-limit estimate $\rho=(4.5\pm0.7)\times10^{-3}$ under the BK18 baseline
deprojection. Deprojecting all six standard modes together with templates for readout
crosstalk from the multiplexing partners and the radially smoothed
counterparts of the standard six
reduces this estimate to $(1.12\pm0.06)\times10^{-3}$. The cross-spectrum
values are \qtylist{83;92}{\percent} of the corresponding auto-spectrum values,
indicating that most of the simulated leakage is reproduced across independent
beam-mapping epochs. Propagating these simulated single-band spectra through
the multi-frequency likelihood determines the
corresponding bias $\Dr$.\cite{verges2026} Crossing the simulated leakage maps
with the real polarization maps instead tests whether the predicted leakage is
present in the data; this comparison awaits unblinding.

The deprojection subspace need not be defined by a parametric beam model. A
Gaussian fiducial gives the familiar correspondence between the standard
differential operators and beam parameters. A finite, tapered aperture instead
has distinct illumination and aperture scales, allowing additional radial
profiles at fixed azimuthal order. The forward model relates these profiles to
first-order perturbations of instrument parameters, while the near-field
comparison supports an association between the wide dipole and detector-pair
differences in offset aperture illumination.

The physical-optics construction can define the wide modes directly, setting
their radial scales from each receiver's modeled optics and providing a common
prescription across BICEP3, BICEP Array, and future BICEP receivers. This connection to instrument parameters may
also allow beam-systematics requirements to better inform future optical design.

The results presented in this work place BICEP3 on target to meet the BK24 requirement
on differential beam response under the current mitigation. Further refinements to our
deprojection procedure offer an avenue for better systematic control as the
\BK and South Pole Observatory programs push to $\sigma(r)\sim0.001$ and
beyond.

\bibliography{report} 
\bibliographystyle{spiebib-etal} 

\end{document}